\documentclass[preprint2,times,tighten]{aastex6}
\pdfoutput=1 %for arXiv submission

\usepackage{amsmath,amstext}
\newcommand{\angstrom}{\textup{\AA}}
\usepackage[figure,figure*]{hypcap}
\usepackage{newtxmath} %use times font for math

\shorttitle{SFHs of active, transitional and quiescent galaxies at $z\sim2$}
\shortauthors{Zick et al.}

\def\dn{D$_n$4000}
\def\hd{H$\delta_{\mathrm{A}}$}
\def\ha{EW(H$\alpha$)}
\begin{document}

\title{The MOSDEF Survey: Stellar Continuum Spectra and Star Formation Histories of Active, Transitional, and Quiescent Galaxies at 1.4$<$\lowercase{z}$<$2.6}

\altaffiltext{1}{Astronomy Department, University of California, Berkeley, CA 94720, USA}

\altaffiltext{2}{Lawrence Livermore National Laboratory, PO Box 808 L-210, Livermore, CA, 94551, USA}
\altaffiltext{3}{Department of Physics \& Astronomy, University of California, Los Angeles, 430 Portola Plaza, Los Angeles, CA 90095, USA}
 \altaffiltext{4}{Department of Physics \& Astronomy, University of California, Riverside, 900 University Avenue, Riverside, CA 92521, USA}
 \altaffiltext{5}{Center for Astrophysics and Space Sciences, University of California, San Diego, 9500 Gilman Dr., La Jolla, CA 92093-0424, USA}

\author{Tom O. Zick\altaffilmark{1,2}, Mariska Kriek\altaffilmark{1}, Alice E. Shapley  \altaffilmark{3},Naveen A. Reddy \altaffilmark{4}, William R. Freeman \altaffilmark{4},Brian Siana \altaffilmark{4}, Alison L. Coil\altaffilmark{5}}

\author{Mojegan Azadi\altaffilmark{6}}\altaffiltext{6}{Harvard-Smithsonian Center for Astrophysics, 60 Garden Street, Cambridge, MA, 02138, USA}

\author{Guillermo Barro\altaffilmark{7}} \altaffiltext{7}{Department of Phyics, University of the Pacific, 3601 Pacific Ave, Stockton, CA 95211, USA}

\author{Tara Fetherolf\altaffilmark{4}}

\author{Francesca M. Fornasini\altaffilmark{5}} 

\author{Laura de Groot\altaffilmark{8}} \altaffiltext{8}{Department of Physics, The College of Wooster, 1189 Beall Avenue, Wooster, OH 44691, USA}

\author{Gene Leung\altaffilmark{7}}

\author{Bahram Mobasher\altaffilmark{4}}

\author{Sedona H. Price\altaffilmark{9}} \altaffiltext{9}{Max-Planck-Institut f{\"u}r extraterrestrische Physik, Postfach 1312, Garching, 85741, Germany}

\author{Ryan L. Sanders\altaffilmark{3}}

\author{Irene Shivaei\altaffilmark{10}} \altaffiltext{10}{Department of Astronomy/Steward Observatory, 933 North Cherry Ave, Rm N204, Tucson, AZ, 85721-0065, USA}
\email{tzick@berkeley.edu}

\begin{abstract}
Using the MOSDEF rest-frame optical spectroscopic survey, we investigate the star-formation histories (SFHs) of different galaxy types, ranging from actively star forming to quiescent at $1.4\leq~z\leq2.6$. SFHs are constrained utilizing stellar continuum spectroscopy, specifically through a combination of Balmer absorption lines, the 4000\,$\angstrom$ break, and the equivalent width of the H$\alpha$ emission line. To attain a sufficiently high signal-to-noise ratio (S/N) to conduct these measurements we stack spectra of galaxies with similar spectral types, as determined from their rest-frame $U-V$ and $V-J$ colors. We bin the MOSDEF sample into five spectral types, subdividing the quiescent and star-forming bins to better explore galaxies transitioning between the two. 
We constrain the average SFHs for each type, finding that quiescent and transitional galaxies in the MOSDEF sample are dominated by an SFH with an average star-formation timescale of $\tau\sim0.1-0.2$~Gyr. These findings contrast with measurements from the low-redshift Universe where, on average, galaxies form their stars over a more extended time period ($\tau>1$~Gyr). Furthermore, our spectral index measurements correlate with mass surface density for all spectral types. Finally, we compare the average properties of the galaxies in our transitional bins to investigate possible paths to quiescence, and speculate on the viability of a dusty post-starburst phase.  

\end{abstract}

\keywords{galaxies: high-redshift, galaxies: evolution}

\section{Introduction}
%%%%%%%%%
A bimodal distribution of galaxy properties (e.g., color, age, morphology) has been observed both in the local universe \citep[e.g.,][]{kauffmann2003} and up to high redshift \citep[e.g.,][]{williams2009,whitaker2011}, defining a red quiescent sequence and a star-forming sequence in color-mass or color-color space. Though a red sequence has been observed out to $z=3$, the relative abundances of quiescent and star-forming galaxies changes across cosmic time; at low redshifts higher mass galaxies are predominantly quiescent, whereas at $z\geq2.5$ star-forming galaxies dominate at all masses \citep[e.g.,][]{muzzin2013,tomczak2014}. However, the process by which these star-forming galaxies quench and join the quiescent sequence remains poorly understood.

\begin{figure*}[t!]
\centering
\includegraphics[width = 6.8in]{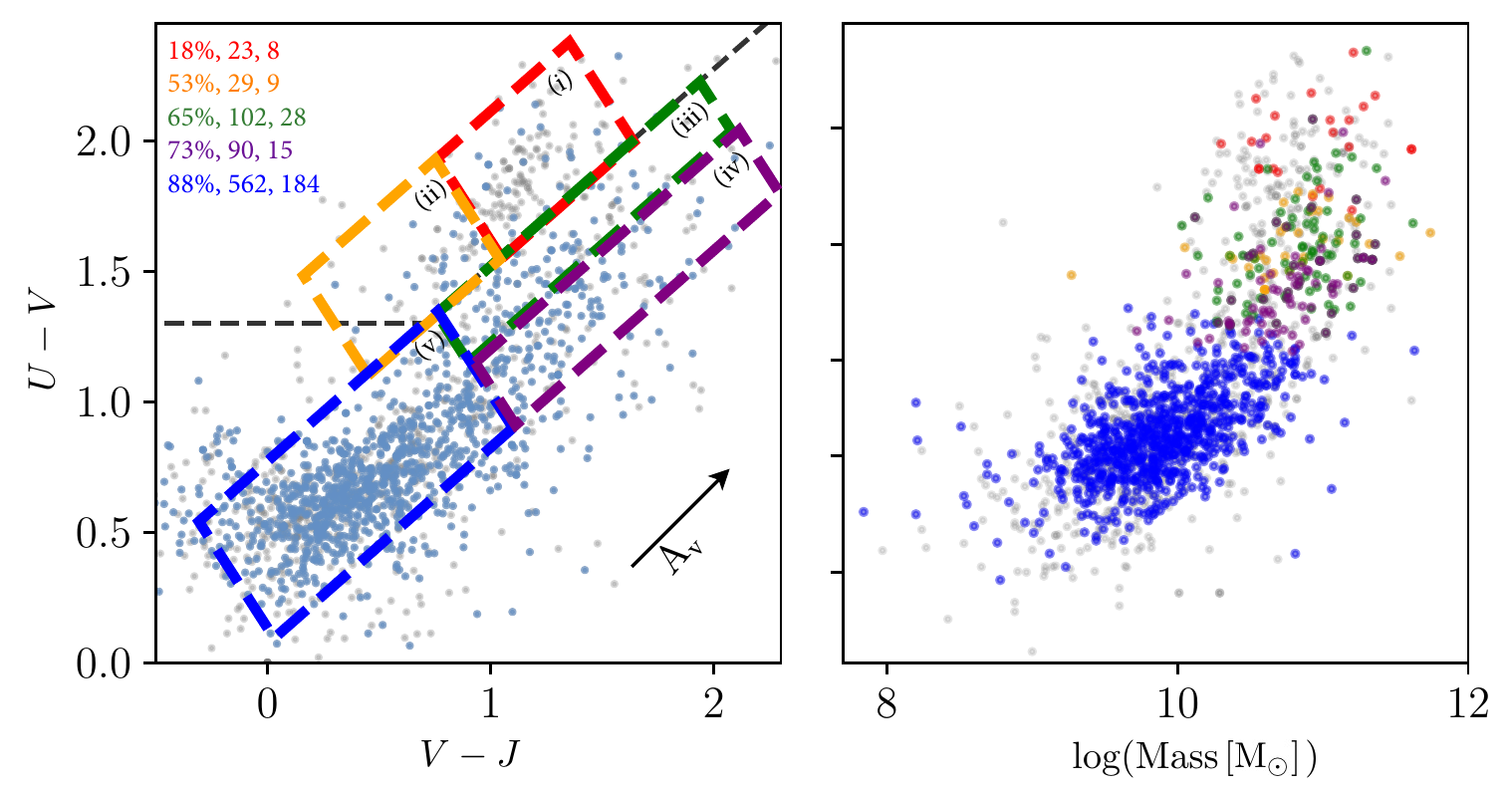}
\caption{Left: The full MOSDEF sample (grey) and our selected sample (blue) in $UVJ$ space, where each box color corresponds to a distinct typical SED shape. On the top left we detail the percent of targets with MOSDEF redshift measurements, the number of spectra with a MOSDEF redshift, and the average number of galaxies per pixel in our composite spectra, colored by corresponding bin. Right: $U-V$ color vs. stellar mass for galaxies (colored by bin) that comprise each stack. }
\label{fig:uvj}
\end{figure*}

Understanding the evolution of galaxies from star forming to quiescent requires detailed knowledge of star-formation histories (SFHs) for a large, representative population of galaxies. While these measurements are readily available at $z\leq0.1$ \citep{kauffmann2003} and have recently been extended to $z\sim0.8$ \citep{wu2018}, the vast majority of quiescent galaxies quench at $z>1$ \citep[e.g.,][]{muzzin2013}. Therefore, understanding quenching requires pushing studies to even higher redshifts. 

Past work around this peak quenching epoch has relied heavily on deep multi-wavelength photometry, which yields poor constraints on SFHs due to model degeneracy, lack of spectroscopic detail, and imprecise redshifts. More robust SFHs can be obtained from the stellar continuum by comparing features sensitive to recent star formation, like the Balmer absorption-line index, \hd, with features sensitive to age, like the 4000~\AA\ break (\dn). Such measurements require high S/N spectroscopy which has in the past only been attained for the brightest, most massive high-redshift sources \citep[e.g.,][]{kriek2009,kriek2016, vandesande2013, belli2015, barro2016}. Alternatively one can stack multiple like galaxies to reach a S/N sufficient to characterize absorption features \citep{onodera2012,mendel2015}, however such studies have also been conducted primarily for massive quiescent galaxies.

With the recently completed MOSFIRE Deep Evolution Field (MOSDEF) survey \citep{kriek2015}, in which MOSFIRE spectroscopy was collected for $\sim1500$ galaxies at $1.37\leq~z\leq3.80$, it is now possible to spectroscopically probe the SFHs of a representative high-redshift galaxy population for the first time. In this Letter we present a technique for constructing composite spectra that conserves stellar continuum, enabling us to measure absorption features sensitive to age and short-term variation in star formation at $z\sim2$. With this methodology, we characterize SFHs for stacks of galaxies across rest-frame $U-V$ vs. $V-J$ color-color space. 

Throughout this work we utilize a \citet{chabrier2003} initial mass function and a $\Lambda$CDM cosmology with $\mathrm{\Omega_M=0.3}$, $\mathrm{\Omega_{\Lambda}=0.7}$ and $\mathrm{H_0=70}$\,$\mathrm{kms^{-1}Mpc^{-1}}$.

\section{Data and Galaxy Sample}

This work leverages the full MOSDEF sample, consisting of rest-frame optical (flux-calibrated) spectra for 1493 H-band selected galaxies between $1.37\leq~z\leq3.80$, with masses and star formation rates (SFRs) ranging from $\sim10^9-10^{11.5}$~M$_\odot$ and $\sim10^0-10^3\,M_\odot$\,yr$^{-1}$, respectively. All MOSDEF galaxies are covered by deep \textit{Hubble Space Telescope}/WFC3 imaging from CANDELS \citep{koekemoer2011,grogin2011}. For information about target selection, data reduction, and sample parameters see \citet{kriek2015}. For the current work we have selected galaxies with a MOSDEF redshift $1.37\leq~z\leq2.61$, as higher redshift galaxies do not typically have sufficient S/N for stacking continuum spectra. We also require at least 400 pixels of coverage within the $\mathrm{3700\,\angstrom\leq\lambda\leq6600\,\angstrom}$ bandpass, which ensures the validity of our stacking method.

For all 806 spectra in our sample we have measured stellar masses and rest-frame colors using the photometric catalogs constructed by the 3D-HST collaboration \citep{skelton2014, momcheva2016} in conjunction with MOSDEF redshifts. Masses are obtained by fitting a galaxy's spectral energy distribution (SED) with stellar population synthesis (SPS) models, utilizing the SPS fitting code \texttt{FAST} \citep{kriek2009} along with flexible SPS models (FSPS) \citep{conroy2009}, and the \citet{calzetti2000} attenuation curve.
We derive rest-frame colors using \texttt{EAzY} \citep{brammer2008} and adopt sizes as measured by \citet{vanderwel2012,vanderwel2014} from the CANDELS/F160W photometric band using \texttt{GalFit} and \texttt{Galapagos} \citep{peng2002,barden2012}. We use the circularized $r_e$ from these size measurements to derive the mass surface density ($\Sigma$) for each of our galaxies.  

\section{Stacking MOSDEF Spectra}

In this work we bin galaxies by similarity in spectral type as determined by their rest-frame $U-V$ vs. $V-J$ colors to attain sufficient S/N per stack to measure SFHs. In the $UVJ$ diagram, dust extinction ($A_v$) increases linearly with increasing $U-V$ and $V-J$, while specific star formation (sSFR) decreases in an almost perpendicular direction for star-forming galaxies. Once a galaxy has stopped forming stars, it will move along the red sequence as it ages \citep[e.g.,][]{wuyts2009, whitaker2012, fumagalli2014, yano2016}. Though galaxies on the quiescent sequence can also be reddened due to increased metallicity, this effect is sub dominant to age \citep{whitaker2013}. We take advantage of these trends in sSFR, age, and dust to bin galaxies as shown in panel (a) of \autoref{fig:uvj}. We use the age gradient in the red sequence to separate our post-starburst (ii) from our quiescent bin (i), and split our star-forming sequence into non-dusty star-forming (v), dusty star-forming (iv) and dusty galaxies with lower sSFRs (iii). We show the bin break down of our selected galaxies in color-mass space in \autoref{fig:uvj}.

For each bin in $UVJ$ space we generate a composite spectrum from individual spectra with varying wavelength coverage. First, we calculate the average best-fit SPS model in luminosity density units for each UVJ bin by averaging the best-fit SPS model per member galaxy. 

Next, for each individual spectrum, we create a skyline mask ($m_{i,x}$) using a S/N cutoff and interpolate the reduced rest-frame spectrum and mask onto a 0.5\,\angstrom\ separated grid to approximate the spectral sampling of MOSFIRE. We derive a scaling parameter ($s_x$) for each individual spectrum using the average best-fit model corresponding to its $UVJ$ bin according to:

\begin{equation}
s_x=\frac{\sum_{i=0}^P\omega_{i,x} r_{i,x}m_{i,x}}{\sum_{i=0}^P\omega_{i,x}f_{i,x}, m_{i,x}}
\end{equation}

\begin{figure*}[th!]
\centering
\includegraphics[width = 6.8in]{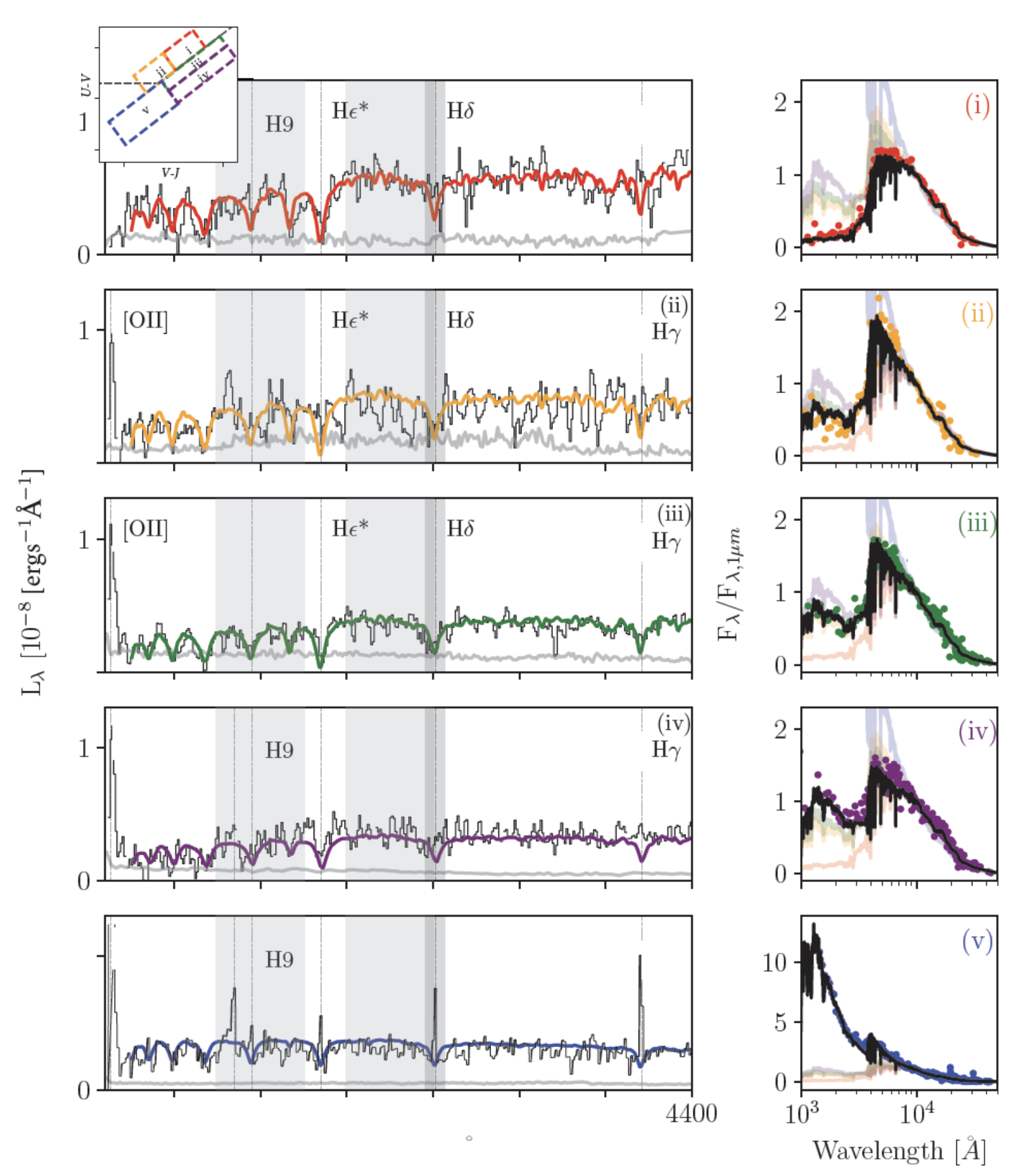}
\caption{\small Left: Stacked spectra for galaxies binned in $UVJ$ space and ordered by rest-frame UV emission relative to 1\,$\mu m$. Each stack (black) and composite noise spectrum (grey) is shown median binned by 2.5\,\AA. The colored lines are the \texttt{FAST} fits to the stacked spectra. The bandpasses from which we measure \dn\ and \hd\ are shown in grey and dark grey respectively. Right: Composite SEDs for each $UVJ$ bin. The colored circles correspond to binned photometric measurements, the black line shows the best fit to the composite SED, while the best fits for the other bins are plotted according to their respective bin color.}

\label{fig:stacks}
\end{figure*}
where P is the total number of pixels in the spectrum, $f_{i,x}$ is the flux of the $i$th pixel of a given spectrum $x$, $\omega_{i,x}$ is its corresponding inverse variance, $r_{i,x}$ is the corresponding luminosity density of the average best-fit SPS model for each bin.

Finally, we stack each spectrum in a given $UVJ$ bin according to a S/N weighted mean stacking method described by:
\begin{equation}
t_i=\frac{\sum_{x=0}^Nw_xs_xf_{i,x} m_{i,x}}{\sum_{x=0}^Nw_xm_{i,x}}
\end{equation}
where $t_i$ is the final stacked value at each pixel and $w_x$ is the average S/N per spectrum. 

By scaling spectra to the average best-fit model for their bin, we correct for flux variations due to redshift differences in our sample and mitigate relative calibration issues between bands for a given galaxy.

\begin{figure*}
\hfill\includegraphics[width=.98\textwidth]{{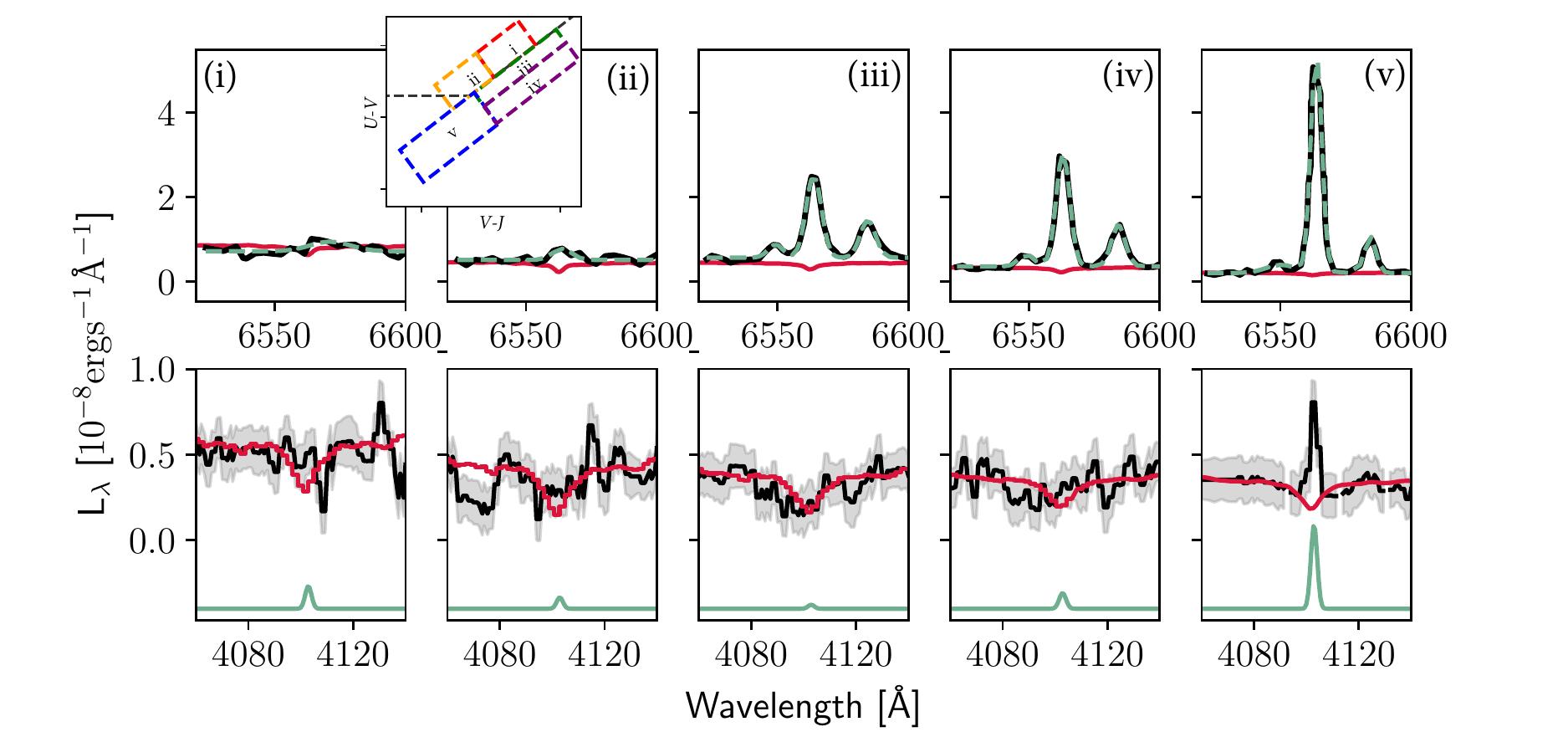}}
\caption{Top: The $\mathrm{H\alpha}$ region for each stacked spectrum (black). The triple Gaussian fit for the spectra is shown in the dashed teal lines, the best fit \texttt{FAST} model is shown in red. Bottom: Zoom in of the \hd\ region, with the best fit for the absorption shown in red and the best fit for the emission (fixed to the H$\alpha$ width), shown in teal. The noise spectrum is also plotted (grey). The emission H$\delta$ in (i) is most likely due to AGN activity, as we only remove AGN from our \ha\ measurements.}
\label{fig:ha}
\end{figure*}

In \autoref{fig:stacks}, we show the resulting composite spectrum and SED for each bin in $UVJ$ space arranged by increasing UV emission relative to the flux at rest-frame 5000\,\AA. In \autoref{fig:ha} we zoom in on the region around H$\alpha$ and H$\delta$ for each of the stacks. 

\section{Measuring Spectral Features}
To measure spectral features for each type, we first mask emission lines from the stacked spectra, then fit them using \texttt{FAST} with high-resolution \citet{bc2003} models. In order to determine the H$\alpha$ emission equivalent width (\ha), we fit with a triple Gaussian to account for potential contamination by the neighboring [N\,{\sc ii}] lines, while correcting for underlying stellar absorption using the best-fit SPS model to the spectrum. We do not apply an additional dust correction to the \ha\ measurements to account for the possibility of increased extinction towards HII regions, thus they may be underestimated. We remove active galactic nuclei from our sample for measurements of H$\alpha$ \citep[see][]{azadi2017}. Next, we simultaneously fit the other Balmer lines in the raw stacks for absorption and emission, fixing the latter to our measured H$\alpha$ linewidths.  
As can be seen in panel (v) of \autoref{fig:stacks}, the emission line widths are considerably narrower than the (pressure broadened) underlying Balmer absorption lines. Given these differing linewidths, we can robustly disentangle the emission from the absorption lines. Finally, we use our emission line fit to subtract the emission lines from the spectrum and re-fit our stacks to measure best-fit star formation timescales. 

We measure continuum features from the emission-line subtracted stacked spectra,
adopting the bandpasses for \hd\ and \dn\ described in \citet{worthey1997} and \citet{balogh1999}, respectively (see \autoref{table:1}). The one exception is the post-starburst stack (ii), where the blue continuum bandpass does not consist of sufficient galaxies to reliably measure \hd\ and we instead measure continuum from the best-fit SPS model to the spectrum.

We derive measurement errors by randomly generating a spectrum, drawing from the noise spectrum of each stack, and repeating our measurements 10000 times. We take the standard deviation of these measurements as our error. We also test for the sensitivity of each bin to its components by repeating our continuum measurements for stacks with a random $\sim10\%$ of each bin removed. Our measurement errors exceed the resulting variation for all but the quiescent stack, which is dominated by two especially massive galaxies. We therefore adopt the standard deviation of the bootstrapped ensemble for the quiescent bin measurements.

\begin{table}
\centering
\begin{tabular}{||c c c c c ||} 
 \hline
 Bin & Log $\tau$ & \hd\  & \dn\ & \ha\\ [0.5ex] 
  & [yr] & [\AA] & & {\centering\arraybackslash [\AA] }\\
 \hline\hline
 i & 8.2 & 4.9 $\pm$  1.1 & 1.44 $\pm$ 0.05 & {\raggedright\arraybackslash}1 $\pm$ 15 \\ 
 ii & 8.4 & 6.2 $\pm$ 3.0 & 1.35 $\pm$ 0.10& {\raggedleft\arraybackslash} 8 $\pm$ 25  \\
 iii & 8.2 & 8.4 $\pm$ 0.4 & 1.29 $\pm$ 0.09& {\raggedleft\arraybackslash}  33 $\pm$ 15 \\
 iv & 9.3 & 6.3 $\pm$ 0.2 & 1.24 $\pm$ 0.09& {\raggedleft\arraybackslash}  56 $\pm$ 10\\
 v & 9.5 & 7.5 $\pm$ 1.3 & 1.14 $\pm$ 0.04& {\raggedleft\arraybackslash}  186 $\pm$ 19\\ [1ex] 
 \hline
\end{tabular}
\caption{Spectral index measurements for each of our stacks along with the best-fit $\tau$ from our \texttt{FAST} fit (SFH of the form: SFR$(t)\propto te^{-t/\tau}$). Each bin number corresponds to $UVJ$ box as shown in \autoref{fig:uvj}.}
\label{table:1}
\end{table}
%Finally, we measure the average reddening (E(B-V)) for our stacks using the Balmer decrement and derive the average E(B-V) per bin using the method described in \citet{reddy2015}.

\section{SFH of \lowercase{z} $\sim2$ Galaxies across $UVJ$ Space}

\begin{figure*}
\centering

\includegraphics[width=6.8in]{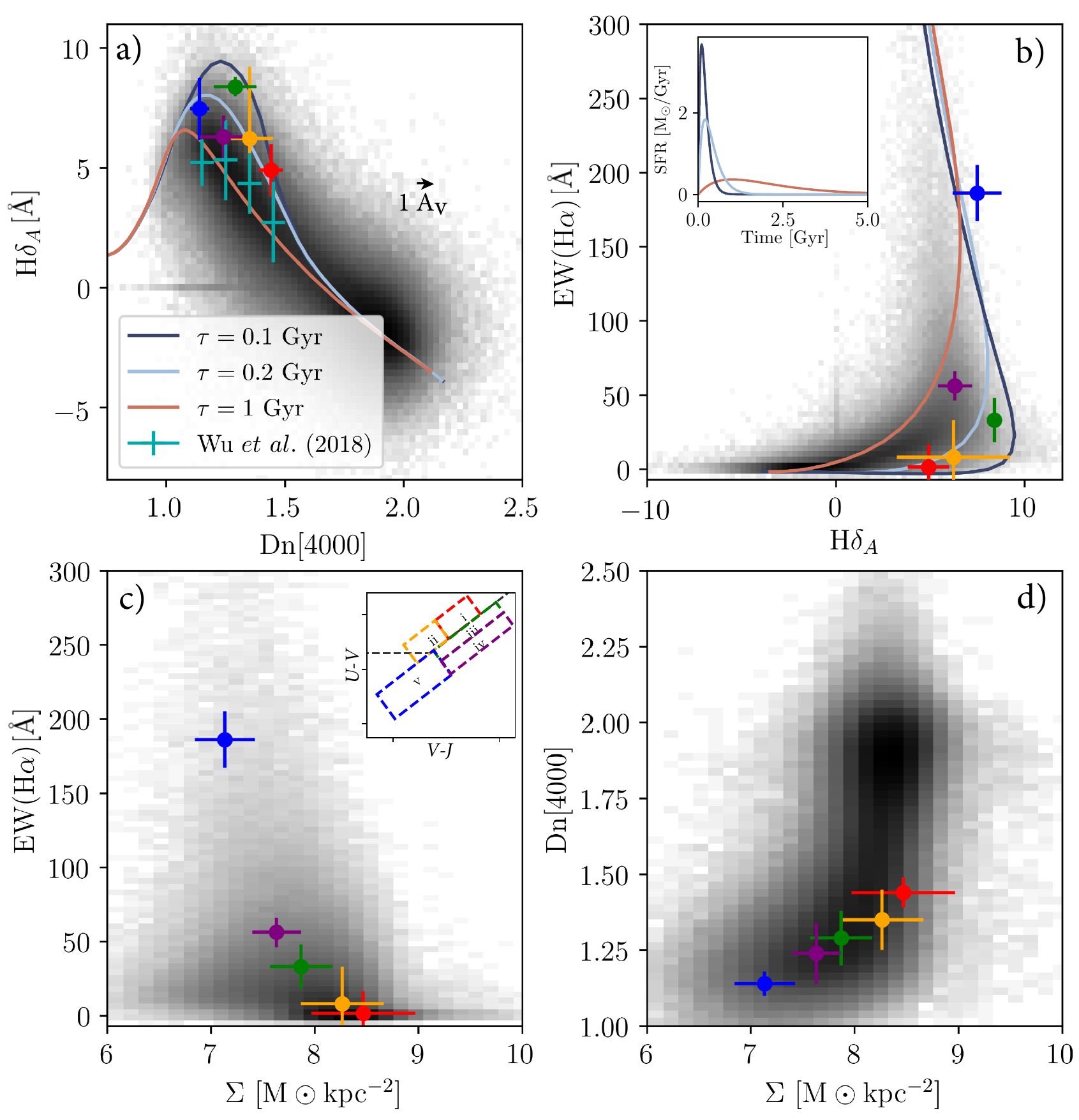}

\caption{\hd, \dn, \ha, and $\Sigma$ measurements for our stacks at $z\sim2$ colored by their respective $UVJ$ bin, compared to low-redshift values from SDSS, shown in greyscale. Where applicable, we overplot the FSPS model tracks for three delayed exponential SFHs with $\tau=0.1,~0.2,~1.0$~Gyr. Panel a: \hd\ vs. \dn, finding higher \hd\ than at low redshift for fixed values of \dn. We also show the $z=0.8$ \citet{wu2018} distribution where the error bars in \dn\ and \hd correspond to bin size and standard deviation of the distribution, respectively. As our \dn\ measurements are not corrected for reddening, we illustrate the effect of 1 A$_V$ of extinction with an arrow. Panel b: \ha\ vs. \hd\ measurements confirming that the star formation timescales of transitional galaxies are most consistent with a short $\tau=100-200$~Myr SFHs. Panels c-d: \ha\ and \dn\ vs. $\Sigma$, illustrating a sequence in decreasing \ha\ and increasing \dn\ as a function of $\Sigma$. Error bars in $\Sigma$ correspond to the standard deviation of the galaxy ensemble in each bin.} 
\label{fig:analysis}
\end{figure*}

We use the spectral features measured in the previous section to infer SFHs for a diverse galaxy population and provide unique insights into galaxy evolution at $z\sim2$. In both Figures \ref{fig:stacks} and \ref{fig:ha} we see clear trends in spectroscopic properties with increasing UV emission relative to rest-frame optical. To asses these trends, we compare \hd, \dn, and \ha\ measured from our stacks in \autoref{fig:analysis}. In panel (a) we compare \hd\ with \dn. The former peaks when A-type stars dominate the spectrum, which only occurs when a relatively short star-formation period is followed by rapid quenching. The latter is sensitive to the opacity of stellar atmospheres and increases with age and metallicity. Comparing the two parameters allows us to assess the evolutionary phase and star-formation time scale of a given galaxy \citep{kauffmann2003}. 

Panel (b) compares \hd\ to \ha, another SFH probe that measures the relative importance of H$\alpha$ emission to the underlying stellar continuum. As continuum emission is a proxy for mass and H$\alpha$ arises from recombination around hot and massive O and early B type stars with short lifetimes, \ha\ relates star formation within the past $50$~Myr to past star formation.

We find that our measurements for the transitional (ii and iii) and quiescent (i) bins form a sequence in both \hd/\dn\ space and \ha/\hd\ space. These bins are most consistent with a rapid star formation model ($\tau=0.1-0.2$~Gyr) shown in panels (a) and (b) of \autoref{fig:analysis} and are therefore most compatible with a fast-quenching SFH. We find consistent results from our best-fit SPS models (see \autoref{table:1}). Additionally, one of our contributing quiescent galaxies has an independent measurement of $\tau\sim200$~Myr from [$\alpha$/Fe] \citep{kriek2016} supporting our measurement. The average SFHs of the galaxies in the dusty (iv) and less-obscured (v) star-forming bins are consistent with a more extended star-formation timescale (i.e., delayed $\tau$ model with $\tau\approx0.2-1$~Gyr). This result supports past work finding constant or rising SFHs for star-forming galaxies at $z=2$ \citep[e.g.][]{maraston2010, lee2010, reddy2012}
\\
We compare our measurements to values from the MPA/JHU SDSS catalogs \citep{kauffmann2003, brinchmann2004} and find an offset between our transitional and quiescent galaxy bins relative to the median sequence at low redshifts. The inferred star-formation timescales in these bins are shorter than for a typical SDSS galaxy ($\tau\geq1$~Gyr) at the $4\sigma$ level, for all but bin (ii), which is significant to 2$\sigma$. This offset is expected, as galaxies in the $z\sim0.1$ universe had a longer period over which stars could have been formed. Nonetheless, the timescales of the transitional galaxies at $z\sim2$ are substantially shorter than the age of the universe at that time. The Lega-C survey also finds an offset at $z=0.8$ for \dn\ vs. \hd\ measured from individual spectra \citep{wu2018}. However, as shown in \autoref{fig:analysis}, it is less pronounced than for our stacks, \textbf{implying shorter star-formation timescales with increasing redshift}.

Past work has shown that quiescent galaxies have a higher mass surface density ($\Sigma$) than star-forming ones \citep[e.g.,][]{barro2014,vandokkum2015}. Mass has also been found to correlate with \dn\, but to a lesser extent \citep[e.g.,][]{kauffmann2003}. As our sample is incomplete in mass in the quiescent region of $UVJ$ space,  but $\Sigma$ is approximately constant within an SED type, we only examine the latter parameter in this work. Using our subdivided blue star-forming and red quiescent bins we assess how $\Sigma$ varies with spectral type. In Figures 4c and 4d, we show \ha\ and \dn\ as a function of $\Sigma$ for each of our stacks. 
We find that \ha\ decreases with increasing $\Sigma$, which -- due to the relative uniformity of mass in all galaxies but those in region (v) of \autoref{fig:uvj} -- is likely primarily due to decreasing SFR with $\Sigma$ and not increasing stellar continuum. We also find that \dn\ increases as a function of $\Sigma$. Taken together, these trends motivate a correlation between decreasing sSFR, increasing age, and $\Sigma$. This relation may be causal as suggested in \citet{vandokkum2015}, or driven by some alternate physical mechanism with which both parameters are correlated \citep{lilly16}. Furthermore, as halos, and by extension star formation, were denser at early times, this sequence could simply be a consequence of the different times at which each of our bins formed their stellar populations with respect to the age of the Universe \citep[e.g.,][]{khochfar2006,abramson18}.

Interestingly, a comparison of our composite spectra and measurements for bins (ii) and  (iii) in \autoref{fig:uvj} implies that these regions contain related stellar populations. Box (ii) corresponds to galaxies characterized by a recent rapid burst of star formation \citep[e.g.,][]{whitaker2012,wild2016} while box (iii) corresponds to what are usually considered to be dusty star-forming galaxies \citep[e.g.,][]{spitler2014}. However, the similarity in \dn\ and \hd\ indicates both regions are comparable in age and have stellar populations dominated by a recent burst of star formation. Comparing \ha\ to \hd\ measurements for (iii), indicates suppressed SFR relative to the rapid past star formation that set \hd. Additionally, galaxies in region (iii) are dusty, with $\mathrm{E(B-V)}=0.41$ derived from the Balmer decrement as described in \citet{reddy2015}. The characteristics of region (iii) galaxies described above, lead us to speculate that these may be dusty post-starburst galaxies that have not yet expelled or depleted their gas and dust reservoirs \citep{poggianti2009}. In this picture, the galaxies in region (iii) could be the progenitors of galaxies found in region (ii) at later times. 

The short star formation timescale we measure for typical galaxies in region (iii) is inconsistent with the picture of a gradual quenching route to quiescence for our redshift range. This result is in contrast with \citet{belli2015}, who examine spectral fitting derived ages and sizes within the quiescent box and suggest the galaxies found in our region (iii) \textbf{may be} progenitors of quiescent galaxies that formed their stellar population over an extended time period, skipping the post-starburst phase all together. It is unclear how to reconcile these results, but the tension may be primarily due to the differing redshift regimes probed in each work; the current study targets slightly higher redshifts. Further measurements may be necessary to understand these discrepancies.

\section{Discussion}

In this letter we constrain SFH as a function of spectral type for a sample of 806 galaxies from the MOSDEF survey at $1.4\leq z\leq2.6$. In order to attain the S/N necessary to constrain SFH from stellar continuum features, we bin galaxies based off of physical trends in $U-V$ vs. $V-J$ and utilize a weighted composite stacking method. We find that transitional and quenched galaxies at $z\sim2$ have a higher \hd\ for a given \dn\ than $z\leq0.8$ galaxies. Specifically, our $z\sim2$ galaxies are consistent with shorter star-formation timescales (100-200\,Myr) as compared to $z<0.8$ ($\geq1$~Gyr). We find a sequence in (increasing) \dn\ and (decreasing) \ha\ with $\Sigma$, highlighting a relationship between evolutionary phase and $\Sigma$, whether it be causal or due to mutual correlation with some third physical parameter. Lastly similarities between the age and SFH of what is usually thought of as part of the dusty star-forming region of the $UVJ$ diagram, to the post-starburst values, motivates that the former may be dusty post-starburst galaxies.    

The unique MOSDEF dataset enabled several improvements compared to past studies. Most importantly, it increased the range of galaxy types for which we could constrain stellar populations and SFHs from stellar continuum spectroscopy, ranging from star forming to transitional to quiescent. Past studies at $z\sim2$ encompassing the full range of galaxy types focused solely on photometric data, while spectroscopic studies based on stellar continuum at comparable redshifts focused on massive quiescent galaxies. Furthermore, stacking galaxies without subtracting a polynomial fit to the continuum as done in previous works, allows us to measure \dn\ for a statistical sample of galaxies for the first time at $z\sim2$. 

Nonetheless, there are several caveats to the current work. First, the MOSDEF survey is slightly biased towards unobscured, star-forming galaxies. This bias, combined with lower success rates for the quiescent (i) and post-starburst (ii) galaxies in the MOSDEF survey, results in a small number of galaxies for these bins. Second, our weighting scheme biases our analysis slightly towards brighter galaxies, as these tend to have higher S/N. This bias primarily affects the bins in which there are few galaxies. Finally, we only consider galaxies with MOSDEF redshifts in this work, so we are biased towards post-starburst and quiescent galaxies with emission lines (from AGN) or bright continuum emission. Future work with NIRSpec on the \textit{James Webb Space Telescope} may overcome these problems without relying on stacking, however such observations at $z\sim2$ will remain challenging. Such larger and more complete samples would enable the use of number density to longitudinally study the evolution of each bin across redshift.

\acknowledgments
We would like to thank Dan Kelson for a referee report that improved the clarity and quality of the manuscript. Funding for the MOSDEF survey is provided by NSF-AAG grants AST-1312780, 1312547, 1312764, and 1313171, grant AR-13907 provided by NASA through a grant from the Space Telescope Science Institute, and NASA-ADAP grant NNX16AF54G. This work was performed under the auspices of the U.S. Department of Energy by Lawrence Livermore National Laboratory under Contract DE-AC52-07NA27344. Funding for TOZ was provided by LLNL Livermore Graduate Scholar Program. We also acknowledge the members of the 3D-HST collaboration, who provided us with spectroscopic and photometric catalogs used to select MOSDEF targets and derive stellar population parameters.
\bibliographystyle{yahapj}
\bibliography{ref.bib}

\end{document}